\documentclass{article}
\usepackage{amsmath,amssymb}
\usepackage{caption}
\usepackage{url}
\usepackage{graphicx}
\usepackage{authblk}

\usepackage{color}
\usepackage{soul}
\soulregister\cite7
\soulregister\ref7
\soulregister\pageref7

\title{
Two Types of Social Grooming Methods depending on the Trade-off between the Number and Strength of Social Relationships
}

\author[1,$\ast$]{Masanori Takano}
\affil[1]{CyberAgent, Inc.}
\affil[$\ast$]{{\small E-mail:~takano\_masanori@cyberagent.co.jp.}}
\date{}

\begin{document}

\maketitle

\begin{abstract}
Humans use various social bonding methods known as social grooming, e.g. face to face communication, greetings, phone, and social networking sites (SNS). SNS have drastically decreased time and distance constraints of social grooming. In this paper, I show that two types of social grooming (elaborate social grooming and lightweight social grooming) were discovered in a model constructed by thirteen communication data-sets including face to face, SNS, and Chacma baboons. The separation of social grooming methods is caused by a difference in the trade-off between the number and strength of social relationships. The trade-off of elaborate social grooming is weaker than the trade-off of lightweight social grooming. On the other hand, the time and effort of elaborate methods are higher than lightweight methods. Additionally, my model connects social grooming behaviour and social relationship forms with these trade-offs. By analyzing the model, I show that individuals tend to use elaborate social grooming to reinforce a few close relationships (e.g. face to face and Chacma baboons). In contrast, people tend to use lightweight social grooming to maintain many weak relationships (e.g. SNS). Humans with lightweight methods who live in significantly complex societies use various social grooming to effectively construct social relationships.
\end{abstract}

\section*{Introduction}


The behaviour of constructing social relationships is called ``social grooming,'' which is not limited to humans but widely observed in primates~\cite{Kobayashi1997,Kobayashi2011,Dunbar,Dunbar2000,Nakamura2003,Sick2014,takano_ngc,takano_socinfo,takano2016_palg}.
Humans use different social grooming methods according to their strength of social relationships~\cite{Burke2014,takano2016_palg} (see also Fig.~1 on electronic supplementary materials (ESM)), e.g. primitive methods (face to face communications) and modern methods (E-mails and social networking sites (SNS)).
Social grooming gives different impressions and has different effects on its recipients depending on the time and effort involved~\cite{Dunbar2012b}.
Face to face communication and on video calls get more satisfaction than communication in phone and text~\cite{Vlahovic2012}.
On Facebook, personal messages give more happiness than 1-click messages (like) and broadcast messages~\cite{Burke2016}.
In other words, humans favor social grooming by elaborate methods (time-consuming and space constrained).
Additionally, people in a close relationship tend to do these elaborate methods~\cite{Burke2014}.
Furthermore, its positive effect in close relationships is larger than in weak social relationships~\cite{Burke2016}.

Humans face cognitive constraints~\cite{Dunbar2012}
(for example, memory and processing capacity) and time constraints (that is, time costs) in constructing and maintaining social relationships.
These time costs are not negligible, as humans spend a fifth of their day in social grooming~\cite{Dunbar1998a} and maintaining social relationships~\cite{Hill2003,ROBERTS2011}.
Therefore, the mean strength of existing social relationships has a negative correlation with the number of social relationships~\cite{Roberts2009,Miritello2013,takano2016_palg}.
The trade-off between the number $N$ and mean strength $m$ of social relationships on online communications (SNS, mobile phones, and SMS) is described as $C=Nm^a$ where $a>1$~\cite{takano2016_palg}, i.e. total communication cost $C$ obeys $Nm^a$.
$C$ which represents the amount of investment for social relationships differs depending on individuals.
This suggests that social grooming behaviour depends on the strength of social relationships and the strength of this trade-off~($a$).

Humans construct and maintain diverse social relationships within the constraints of this trade-off.
These relationships provide various advantages to them in complex societies.
Close social relationships lead to mutual cooperation~\cite{Harrison2011,haan2006,takano_ngc,takano_socinfo}.
On the other hand, having many weak social relationships, i.e. weak ties, help in obtaining information, which is advantageous because weak social relationships where people rarely share knowledge often provide novel information~\cite{Granovetter1973,Dunbar,Onnela2007,Ellison2007,Eagle2010}.

As a result, social relationship forms (distributions of social relationship strengths) often show a much skewed distribution~\cite{Zhou2005,Arnaboldi2013b} (distributions following a power law~\cite{Hossmann2011,Fujihara2014,takano2016_palg}).
This skewed distribution has several hierarchies called circles.
The sizes of these circles (the number of social relationships on the inside of each circle) are 5, 15, 50, 150, 500, and 1500, respectively~\cite{Zhou2005,Dunbar2018}.
That is, the ratios between neighboring circles of social relationships are roughly three irrespective of social grooming methods, e.g. face to face, phone, Facebook, and Twitter~\cite{Dunbar2018}.

I aim to explore how and why humans use various social grooming methods and how those methods affect human behaviour and social relationship forms.
For this purpose, I analyze the strength of the trade-off between the number and mean strength of social relationships as a key feature of social grooming methods.
For this analysis, I extend the model~\cite{takano2016_palg} for $C=Nm^a$ from $a>1$ to $a>0$ by introducing individuals' strategies about the amount of social grooming behaviour.
This model explains not only online communication but also offline communication.
The key feature restricts social relationship forms (size and distributions of strengths).
Therefore, humans should change social grooming strategies depending on the trade-off, i.e. they tend to use several social grooming methods for constructing various strengths of social relationships.
My model is supported by the common features of thirteen diverse communication data-sets including {\it primitive human communication} (face to face and communication in a small community constructed by kin and friends), {\it modern communication} (phone calls, E-mail, SNS, and communication between unrelated people), and {\it non-human primate communication} (Chacma baboons).
The model connects social behaviour and social relationship forms with a trade-off between the number and mean strength of social relationships.

\section*{Data Analysis}

I found two types of social grooming methods based on the trade-off between the number and strength of social relationships (Fig.~1).
One was ``elaborate social grooming,'' which was face to face and by phone (Face to face (Pachur)~\cite{Pachur2014}\footnote{\url{https://doi.org/10.5061/dryad.pc54g}} and Phone (Pachur)~\cite{Pachur2014}\footnote{\url{https://doi.org/10.5061/dryad.pc54g}}), in kin and friends (Mobile phone (friends and family)~\cite{Aharony2011}\footnote{\url{http://realitycommons.media.mit.edu/realitymining4.html}} and Short Message Service (SMS (friends and family))~\cite{Aharony2011}\footnote{\url{http://realitycommons.media.mit.edu/realitymining4.html}}), and Chacma baboon social grooming (Baboon group A and B)~\cite{Sick2014}\footnote{\url{https://doi.org/10.5061/dryad.n4k6p/1}}.
This should be nearer to primitive human communications than the others.
That is, these communications tend to bind individuals due to time and distance constraints or in primitive groups constructed by kin and friends.
Another one was ``lightweight social grooming'' which was by SNS and E-mail (Twitter~\cite{Cheng2010}\footnote{\url{https://archive.org/details/twitter_cikm_2010}}, 755 group chat~\cite{takano2016_palg}\footnote{\url{https://doi.org/10.6084/m9.figshare.3395956.v3}}, 755 wall communication~\cite{takano2016_palg}\footnote{\url{https://doi.org/10.6084/m9.figshare.3395956.v3}}, Ameba Pigg~\cite{takano2016_palg}\footnote{\url{https://doi.org/10.6084/m9.figshare.3395956.v3}}, and E-mail/Letter (Pachur)~\cite{Pachur2014}\footnote{\url{https://doi.org/10.5061/dryad.pc54g}}), and in relationships between unrelated people (Mobile phone (dormitory)~\cite{Madan2012}\footnote{\url{http://realitycommons.media.mit.edu/socialevolution.html}} and SMS (dormitory)~\cite{Madan2012}\footnote{\url{http://realitycommons.media.mit.edu/socialevolution.html}}), which has appeared in the modern age.
These communications tend to unbind humans from time and distance constraints.
This tended to be used with unrelated people.
Details of data-sets noted in brackets are shown in the Data-Sets section on ESM.

Both were divided by parameter~$a$ on $C_i=N_i m_i^a$ model (Fig.~2), where $a$ showed strengths of the trade-off between $N_i$ and $m_i$, and individual~$i$'s total social grooming cost was $C_i$, $N_i$ was $i$'s number of social relationships, $m_i$ was $i$'s mean strength of their social relationships ($m_i = \sum_{j=1}^{N_i} d_{ij} / N_i$), and $d_{ij}$ was the total number of the days on which $i$ did social grooming to individual~$j$ (the strength of social relationships between $i$ and $j$).
$C_i$ represents $i$'s available social capital on each social grooming method which varied widely among individuals (see Fig.~3).
I estimated statistically parameter~$a$ of the data-sets by using a regression model $\log N \sim Normal(-a \log m + b \log u, \sigma)$, where $u$ was the number of days of participation for each person and $\sigma$ was a standard deviation, that is, this model assumed that a user's total social grooming costs were equal to the $b$-th power of the number of days for which they had participated in the activity ($C = u^b$)~\cite{takano2016_palg}.
$u^b$ was entered as a covariate to control the usage frequency of social grooming methods.
Table~1 shows the details of this regression result.
A few p-values of $a$ were larger than $0.05$, but social grooming methods overall seem to be divided by parameter~$a$.

$a \neq 1$ suggests that social grooming behaviour depends on the strength of social relationships because $a$ will be $1$ when social grooming behaviour does not relate to the strength of social relationships ($C_i=N_i m_i =\sum_{j=1}^{N_i} d_{ij}$).
$a<1$ shows that people have stronger social relationships than when $a=1$ because the effect of strong relationships (large $m$) to cost $C$ is smaller than when $a=1$ ($C_i=N_i m_i^a < \sum_{j=1}^{N_i} d_{ij}$ when $m>1$).
In contrast, $a>1$ shows that people have weaker social relationships than when $a=1$ because of the effect of strong relationships (large $m$) to cost $C$ is larger than when $a=1$ ($C_i=N_i m_i^a > \sum_{j=1}^{N_i} d_{ij}$ when $m>1$).
I call the social grooming methods of $a>1$ (i.e. modern methods) lightweight social grooming (Fig.~2a-g) and the methods of $a<1$ (i.e. primitive methods) elaborate social grooming (Fig.~2h-m).
That is, the trade-off of elaborate social grooming between the number and strength of social relationships is smaller than that of lightweight social grooming.
On the other hand, the time and effort of lightweight methods are less than elaborate methods~\cite{Dunbar2012a}.

This trade-off parameter $a$ affected human social grooming behaviour.
People invested more time in their closer social relationships, i.e. the amount of social grooming between individuals increased with their strength of social relationships.
Additionally, $a$ changed people's trends of social relationship constructions.
People having limited and deep social relationships tended to do frequent social grooming (amount of social grooming was large) when $a<1$.
On the other hand, people having expanded and shallow social relationships tended to do frequent social grooming when $a>1$.
I show these in the following.

Fig.~4, 5, and the previous work~\cite{takano2016_palg} show that the amount of social grooming from individual~$i$ to individual~$j$ tended to increase with the density of social grooming $w_{ij}$, where $w_{ij}=d_{ij}/t$ and $t$ was the number of elapsed days from the start of observation, i.e. the amount of social grooming did not depend on $t$.
I modeled this phenomenon as linear increase which was the simplest assumption, that is, $v(w_{ij}) = \alpha w_{ij} + 1$, where $v(w_{ij})$ was the amount of social grooming from individual $i$ to individual $j$ and $\alpha$ was a parameter.

This assumption and the definition of $C_i=N_i m_i^a$ suggest a relationship between individual social relationship trends and their total amount of social grooming.
$m_i$ shows individual~$i$'s sociality trend (mean limitation and depth of its social relationships) by the trade-off between $N$ and $m$.
The total amount of social grooming per day to reinforce social relationships is
\begin{align}
G(a, \alpha; C, m) = \alpha C (m^{1-a}  - m^{-a})/T,
\label{eq_gi}
\end{align}
where $T$ is the number of days of the data periods.
$m_i=N_i^{-1} \sum_{j=1}^{N_i} d_{ij}(T) = T N_i^{-1} \sum_{j=1}^{N_i} w_{ij}$ based on the definition of $m$ and $w$.
Therefore, the total amount of social grooming $V_i$ is $\sum_j^{N_i}v(w_{ij}) = \alpha m_i N_i T^{-1} + N_i$.
$G(a, \alpha; C, m)$ is acquired by subtracting the total cost for creating social relationships ($V_0 = \alpha + T$) from $V_i$ (see the Development of Eq.~1 section on ESM).
$\alpha$ was decided by the following simulation where an individual-based model was fitted to the data-sets.
This equation shows that people who have large $m$ (i.e. limited and deep social relationships) have a large amount of social grooming when doing elaborate methods ($a<1$; the orange line in Fig.~13d).
On the other hand, people who have small $m$ (i.e. expanded and shallow social relationships) have a large amount of social grooming when doing lightweight methods ($a>1$; the green line in Fig.~13d).
That is, $G(a, \alpha; C, m)$ shows that social grooming methods were used depending on the strength of social relationships (elaborate social grooming used for strong social relationships and lightweight social grooming used for weak social relationships).
The threshold of both social grooming methods was $a=1$.
The amount of social grooming for construction of new social relationships $G_0$ does not depend on $a$ ($G_0 \simeq N$; see Development of Eq.~1 section on ESM for details).

$a$ changed the peak of a total amount of social grooming $G$ depending on sociality trend $m$, nevertheless, the number of social relationships for each social grooming method did not show clear relationships to $a$ (Fig.~6).
Additionally, their number of social relationships which were smaller than the general number (about 150~\cite{Pollet2011b,Dunbar2016}). 
There may be the problem for evaluating the number of social relationships due to the differences in data gathering on the data-sets.
There may be not a difference of the number of social relationships among the data-sets because the previous studies~\cite{Pollet2011b,Dunbar2016} showed that the number of social relationships did not particularly depend on social grooming methods.

$a$ affected the ratio of very weak social relationships.
Fig.~7 shows the size ratio between neighboring hierarchies of social relationships, i.e. $H_k/H_{k+1}$, where $H_k$ is the number of social relationships when $d \geq k$.
That is, $H_1$ is the number of all social relationships, $H_2$ is the number of social relationships excluding relationship $d=1$, and $H_k$ with large $k$ is the number of close relationships.
The trade-off parameter $a$ only affected $H_k/H_{k+1}$ when $k$ was very small.
Thus, the strength of trade-off $a$ related only to very weak social relationships which seemed to be the circle of acquaintances~\cite{Dunbar2018}.

\section*{Individual-based Simulations}
\subsection*{Model}

In the previous section, I found the threshold $a=1$ on social grooming behaviour (Eq. 1).
In contrast, a threshold was not observed on social relationship forms due to the differences of data gathering and sampling on the data-sets.
In this section, I conducted individual-based simulations to analyze changes of social relationship forms depending on $a$ under the same conditions.

I constructed an individual-based model to explore the effects of the trade-off parameter~$a$ to social relationship forms based on the monotonic increasing of $v(w_{ij})$ and the difference of the peak of $G(a, \alpha; C_i, m_i)$ depending on $a$ (Eq.~\ref{eq_gi}).
That is, individual $i$ does social grooming to individual $j$, the amount is proportional to $w_{ij}$.
$i$'s total amount of social grooming for reinforcing all social relationships is $G(a, \alpha; C_i, m_i)$
Additionally, I assumed the Yule--Simon process on social grooming partner selection, because people basically do act this way~\cite{Pachur2014,takano2016_palg} (see also ESM Fig.~2).
In the Yule--Simon process, which is one of the generating processes of power law distributions~\cite{Newman2005}, individuals select social grooming partners in proportion to the strength of their social relationships, that is, the individuals reinforce their strong social relationships.
In the model, individuals construct new social relationships and reinforce existing social relationships, where they pay their limited resources $R=G(a, \alpha; C_i, m_i)$ for the reinforcement.
This model is an extension of the individual-based model of the previous study~\cite{takano2016_palg} in which $G$ is introduced.
The Source Code of the Individual-based Simulations section on ESM shows a source code of this model.

I consider two types of individuals, groomers and groomees.
Groomers construct and reinforce social relationships using their limited resources $R$ (that is, time), based on these assumptions and the Yule--Simon process.
I use the linear function $v(w_{ij}) =\alpha w_{ij} + 1$ as the amount of social grooming from groomer $i$ to groomee $j$ as with the above section.

I conducted the following simulation for $T$ days to construct social relationships $d_{ij}$ in simulation experiments 1 and 2.
Individuals have a social relationship where strength is 1 as the initial state.
On each day $t \in [1, T]$, groomer $i$ repeats the following two-processes for its resource $R_i > 0$.
$R_i$ is reset to an initial value $G(a, \alpha; C_i, m_i)$ before each day $t$.
Each $i$ spends $R_i$ reinforcing its social relationships.

\paragraph{Creating new social relationships}
Each $i$ creates social relationships with strangers (groomees).
The strength of a new social relationship with $j$ ($d_{ij}$) is 1.
The number of new relationships obeys a probability distribution $Poisson(p_i)$, where $p_i$ is $(N_i-1)/T$.
Therefore, $i$ is expected that it has $N_i$ social relationships until day $T$,
because the relationship between $N$ and $a$ was unclear in the previous section, the expected value of $N$ does not depend on $a$ and is constant.
This setting should be natural because the previous studies~\cite{Pollet2011,Dunbar2016} showed independence of $N$ from social grooming methods.
Creating new relationships does not spend $R$.

\paragraph{Reinforcing existing social relationships}
$i$ also reinforces its social relationships.
Each $i$ selects a social grooming partner $j$ depending on a probability proportional to the strength of the social relationships between $i$ and $j$, then $i$ adds $1$ to $d_{ij}$ (that is, the Yule--Simon process) and spends the amount of social grooming $v(w_{ij})$ from $R_i$ (if $R_i < v(w_{ij})$, then $i$ adds $R_i/v(w_{ij})$ to $d_{ij}$ and $R_i$ becomes $0$).
Each $i$ does not perform the act of social grooming more than once with the same groomees in each day $t$.
Therefore, selected groomees are excluded from the selection process of a social grooming partner $j$ on each day $t$.

\subsection*{Simulation Experiment 1: Checking the Model Consistency}

In this experiment, I confirmed a consistency between $C=Nm^a$ and two assumptions of social grooming behaviour ($G(a, \alpha; C_i, m_i)$ and $v(w_{ij}) =\alpha w_{ij} + 1$).
Therefore, I fitted my individual-based model to the data-sets optimized by unknown parameter $\alpha$.
I used actual values of the data-sets as $a, T$, and $C_i$ in each simulation, where $a$ was the values in Fig.~2 and Table~1, $T$ was the period for each data-set, and $C$ was the $75$th percentile of $u^b$.
$N_i$ was equally divided in a logarithmic scale ($N \in [1, T]$).
Unknown parameter~$\alpha$ was calculated by the optimization which decreased error values of simulations ($e=\sum_{i=1}^M \{(\log N_i - \log N'_i)^2 + (\log m_i - \log m'_i)^2 \}/ M$), where $m_i = (C/N_i)^{(1/a)}$, $M$ was the number of individuals ($M=30$), and $N'_i$ and $m'_i$ were calculated by simulation results (social relationship strengths $d$ of each individual).

Next, I calculated social relationship forms ($d_{ij}$) in each $a$ by using actual settings ($T, C_i, N_i$) and the optimized $\alpha$, where $T$ was the period for each data-set, $C_i$ was individual $i$'s $u_i^b$, and $N_i$ was $i$'s number of social relationships in each data-set.

This model fit all data-sets (Fig.~8 and Table~2).
Their distributions of social relationships $d_{ij}$ were roughly similar to actual distributions excluding Face to face (Pachur) (Fig.~9).
Additionally, the amount of social grooming predicted by Eq.1 with the optimized $\alpha$ showed a high correlation with the actual amount of social grooming in each data-set (Table 3).
That is, this model roughly has an explanation capacity for generating the process of social relationships, depending on human social grooming behaviour, regarding the trade-off constraint.
The difference between the simulation result and Face to face (Pachur) data-set may have been because the approximations of this model did not work with small $a$.

\subsection*{Simulation Experiment 2: Effects of Social Grooming Methods}

In this experiment, I analyzed the effect of parameter $a$ on the structure of social relationships by using the model.
First, I calculated error value $e_{a\alpha}$ in each $a$ and $\alpha$, where $a$ was $\{0.50, 0.55, \dots, 2.00\}$ and $\alpha$ was $\{1.00, 1.02, \dots, 3.00\}$.
That is, the number of combinations is $31 \times 101 = 3,131$.
In each simulation, $T$ was the period for the Twitter data-set, $C_i$ was the $75$th percentile of $u^b$ in the Twitter data-set, $N_i$ was equally divided in a logarithmic scale, and $M=30$.
Each $e_{a\alpha}$ was calculated fifty times, i.e. there are $5,050$ results on each $a$.
I used $\alpha$ which was ranked in the lowest twenty of $e_{a\alpha}$ of each $a$.

Next, I calculated social relationship forms (the distributions of social relationship strengths $d_{ij}$) in each $a$ by using actual settings ($T, C_i, N_i$) and $\alpha$, where $T$ was the period for the Twitter data-set, $C_i$ was individual $i$'s $u_i^b$, $N_i$ was $i$'s number of social relationships in the Twitter data-set, and $M$ was the number of people in the Twitter data-set.

Firstly, I evaluated a power law coefficient $\phi$ as overall effects of $a$ on social relationship forms because social relationship forms (distributions of social relationship strengths $d_{ij}$) follow power-law distributions~\cite{Hossmann2011,Fujihara2014,takano2016_palg}.
Secondly, I analyzed the ratio between neighboring circles of weak social relationships which depended on $a$ in the previous section (Fig.~7).

I found that $a$ changed the social relationship forms $\phi$ and social behaviour parameter $\alpha$ where the threshold was around $a=0.8$ (Fig.~10) because a linear regression model with the threshold at $a=0.8$ was more accurate than a model without the threshold, i.e. the former had smaller AIC than the latter.
The former is $\phi \sim Normal(\beta_1 a f + \beta_2 a (1 - f) + \beta_3 f + \beta_0, \sigma)$, where $f=1$ when $a \geq 0.8$ otherwise $f=0$ and $\sigma$ is standard deviations (AIC: $-1452.6$; see Table 4).
The latter is $\phi \sim Normal(\beta_1 a + \beta_0, \sigma)$ (AIC: $-1400.9$).
That is, the changes of $a$ in $a<0.8$ ($\beta_2$) had a smaller effect on powerlaw coefficients $\phi$ of social relationship forms than in $a\geq0.8$ ($\beta_1$).
This shows that strong social relationships decreased in $a\geq0.8$ because individuals having expanded and shallow social relationships have more of the amount of social grooming than individuals having limited and deep social relationships ($G(a, \alpha; C, m)$).
This threshold seems to be due to the threshold of $G(a, \alpha; C, m)$.
The difference between this threshold ($a=0.8$) and the threshold of $G(a, \alpha; C_i, m_i)$ ($a=1$) may have been because of the approximations of this model.

Additionally, $\alpha$ in $a<0.8$ was larger than $\alpha$ in $a\geq0.8$ excluding $a=0.8$.
Interestingly, $\alpha$ also drastically changed in the range of $0.8\leq a\leq 1.3$.
That is, individuals in $a\geq0.8$ decreased the amount of social grooming $v(w_{ij})$ with close social relationships as compared to $a<0.8$.

$a$ affected the ratio of very weak social relationships.
Fig.~11 shows the size ratio between neighboring hierarchies of social relationships, i.e. $H_k/H_{k+1}$, where $H_k$ is the number of social relationships when $d > k$.
That is, $H_1$ includes all social relationships excluding relationships of $d \leq 1$ (one time interactions), $H_2$ is the number of social relationships excluding relationship $d \leq 2$, and $H_k$ with large $k$ is the number of close relationships.
There is the difference of the definition between this $H$ and $H$ in the Data Analysis section due to the difference of the definition between $d$ in this section (positive real number) and $d$ in the Data Analysis section (natural number).
The ratio between neighboring hierarchies of very weak social relationships $H_1/H_2$ significantly changed around $a=0.8$.
On the other hand, $H_k/H_{k+1}$ on $k \geq 2$ gradually increased with $a$.
This was due to the fact that $H_1$ increased with the increase $a$ and $H_k$ when $k \geq 2$ decreased with the increase $a$ (Fig.~12).
Thus, very weak social relationship forms were especially affected by $a$ compared with strong social relationships.

As a result, the social relationship forms were expanded and shallow in $a\geq0.8$.
This suggests that societies with lightweight social grooming had different properties when compared to societies with elaborate social grooming.

\section*{Discussion}

I constructed a model of social relationship forms depending on human behaviour restricted by a trade-off between the number and strength of social relationships depending on social grooming methods.
This model was supported by common features of thirteen diverse communication data-sets including primitive human communication, modern communication tools, and non-human primates.
By analyzing the model, I found two types of social grooming (elaborate social grooming and lightweight social grooming).
They made different social relationship forms.
This was caused by people's social grooming behaviour depending on the different trade-off between the number and strength of social relationships.
Both methods were separated by trade-off parameter~$a$ on $C=Nm^a$ model ($a<1$: elaborate social grooming, $a>1$: lightweight social grooming).
This separation was due to the total amount of social grooming $G$ having the threshold $a=1$.
The model from the previous study~\cite{takano2016_palg} was expanded by adding $G$.

People tended to use elaborate social grooming in face to face communication and communication in small communities made up of kin and friends, i.e. the communities should be near primitive groups.
Additionally, Chacma baboons also showed a similar trend.
They tended to use this social grooming to reinforce close social relationships.
That is, elaborate social grooming is a primitive method (i.e. a priori).
This may be used in non-human primates, primitive human societies, and close relationships of modern humans, i.e. these may not have a qualitative difference.

On the other hand, people tended to use lightweight social grooming in SNS, E-mail, and communication in communities made up of unrelated people, i.e. the communities should be non-primitive groups.
That is, this social grooming is posterior.
People tended to use these methods to construct many weak social relationships.
As a result, social relationship forms may have changed significantly when people have used lightweight social grooming.
Therefore, human societies have become expanded and shallow.
Fig.~13 shows the relationship between the trade-off parameter of social grooming methods, human social behaviour, and social relationship forms.

Due to these differences, people use both social grooming methods depending on the strengths of social relationships~\cite{Burke2014,Wohn2017,Kushlev2017} (Fig.~1 on ESM).
A typical person who has various strengths of social relationships~\cite{Zhou2005,Dunbar2018} uses elaborate social grooming ($a<1$) for constructing a few close relationships and lightweight social grooming for having many weak social relationships.
This is caused by the change of the peak of the function $G$ around threshold $a=1$.

The function $G$ represents the total amount of social grooming to all relationships depending on sociality trend $m$ and trade-off parameter $a$ of social grooming.
That is, individuals' total amount of social grooming are limited, and these limits ($G(a, \alpha; C, m)$) depend on $m$ and $a$.
Thus, some individuals opt for lightweight social grooming ($a>1$) a consequence of the fact that they want to have a larger network (small $m$ and large $N$).
In other words, two types of social grooming have emerged in consequence of social grooming strategies, which are how individuals distribute their limited resources by using several social grooming methods depending on trade-off parameter $a$ of the social grooming.

This qualitative difference between the two types of social grooming may be caused by the number of very weak social relationships an individual wants to create.
The most affected relationships according to the difference of social grooming methods were the ratio between a hierarchy of social relationships including very weak social relationships and its neighboring hierarchy.
On elaborate social grooming, the trade-off parameter $a$ has an insignificant effect on this ratio.
In contrast, on lightweight social grooming, an increase of the trade-off parameter $a$ increases the ratio.
Humans may have acquired lightweight social grooming by the necessity to create many very weak social relationships (acquaintances).
It seems to have been due to the increase in the number of accessible others or community sizes.

Some online social grooming methods (Mobile phone (friends \& family), SMS (friends \& family), and Phone (Pachur)) showed $a<1$.
These might have been caused by using for constructing non-weak social relationships.
That is, the subjects of Mobile phone (friends \& family) and SMS (friends \& family) were members of a young family living in a residential community which was constructed by kin and neighbors.
Phone (Pachur) was used for communication closer relationships than E-mails when people use phone and E-mails (ESM Fig.~1).
This could be achieved by comparing diverse communication data-sets gathered by similar conditions.

Both social grooming methods also differ from a cost and effect perspective.
The trade-off of interacting in close social relationships using elaborate methods is weaker than that of lightweight methods.
On the other hand, the time and effort of lightweight methods are less than elaborate methods~\cite{Dunbar2012a}.
Social grooming with time and effort (elaborate social grooming) is effective to construct close social relationships~\cite{Dunbar2012a,Vlahovic2012,Burke2016}.
Therefore, elaborate methods are suited to maintain a few close relationships.
In addition, lightweight methods make it easier for people to have many weak social relationships~\cite{Ellison2007,Arnaboldi2013}.

The two types of social grooming methods have different roles.
The role of elaborate methods should be to get cooperation from others.
Humans tend to cooperate in close friends~\cite{haan2006,takano_socinfo,Harrison2011,Miritello2013a,Dunbar2016,Sutcliffe2016} because cooperators cannot cooperate with everyone~\cite{Xu2015,Sutcliffe2016}.
The role of lightweight social grooming should be to get information from others.
Weak social relationships tend to provide novel information~\cite{Granovetter1973,Dunbar,Ellison2007}.

Thus, it should be effective for people to use elaborate social grooming to close relationships while expecting cooperation from these relationships.
They use widely lightweight social grooming in weak relationships while expecting novel information.
As a result, the number of close relationships before and after SNS has not changed much~\cite{Dunbar2016}.
Weak relationships after the appearance of SNS have been maintained effectively~\cite{Burke2014,takano2016_palg}.

An advantage of having information would have increased with the changes of societies.
As a result, lightweight social grooming has been necessary, and humans have had expanded and shallow social relationship forms.
Humans probably have acquired this social grooming in the immediate past.
This consideration will become clearer by analyzing various data-sets, e.g. other non-human primates, social relationship forms in various times and cultures, and other communication systems.

The way of using both methods may also depend on people's extroversion/introversion.
In general, introverts have limited deep social relationships and extroverts have expanded shallow social relationships~\cite{Roberts2009a,Pollet2011a}.
The diversity of the amount of social grooming $C$ on each social grooming method suggests that usage strategies of the two types of social grooming methods differ for each person, e.g. introverts tend to do elaborate social grooming, in contrast, extroverts tend to do lightweight social grooming.

Primitive humans also used lower-cost social grooming methods (e.g. gaze grooming~\cite{Kobayashi1997,Kobayashi2011} and gossip~\cite{Dunbar}) than fur cleaning in non-human primates.
These methods have evolved more for larger groups than that of non-human primates because these grooming methods enable humans to have several social relationships and require less time and effort~\cite{Dunbar1998,Dunbar2000}.
However, my model does not distinguish these social grooming methods from social grooming in non-human primates; nevertheless, the model separates modern social grooming from primitive human social grooming.
This may suggest that an appearance of lightweight social grooming significantly affects human societies nearly as much as the changes between non-human primates and primitive humans.

\section*{Data Accessibility}
All data needed to evaluate the conclusions in the paper are present in the paper and the electronic supplementary materials.


\section*{Competing Interests}
Masanori Takano is an employee of CyberAgent, Inc. There are no patents, products in development or marketed products to declare.
This does not alter the authors' adherence to all Royal Society Open Science policies on sharing data and materials, as detailed online in the guide for authors.

\section*{Funding}
I received no funding for this study.

\section*{Research Ethics}

This study did not conduct human experiments (all data-sets were published by previous studies).

\section*{Animal Ethics}

This study did not conduct animal experiments (all data-sets were published by previous studies).


\section*{Permission to Carry out Fieldwork}

This study did not conduct fieldwork.

\section*{Acknowledgements}
I are grateful to associate professor Genki Ichinose at Shizuoka University, Dr. Vipavee Trivittayasil at CyberAgent, Inc., Dr. Takuro Kazumi at CyberAgent, Inc., Mr. Hitoshi Tsuda at CyberAgent, Inc. and Dr. Soichiro Morishita at CyberAgent, Inc. whose comments and suggestions were very valuable throughout this study.


\clearpage
\footnotesize

\begin{figure*}[ht!]
\begin{center}
\captionsetup{width=1.4\linewidth}
\includegraphics[width=.6\linewidth]{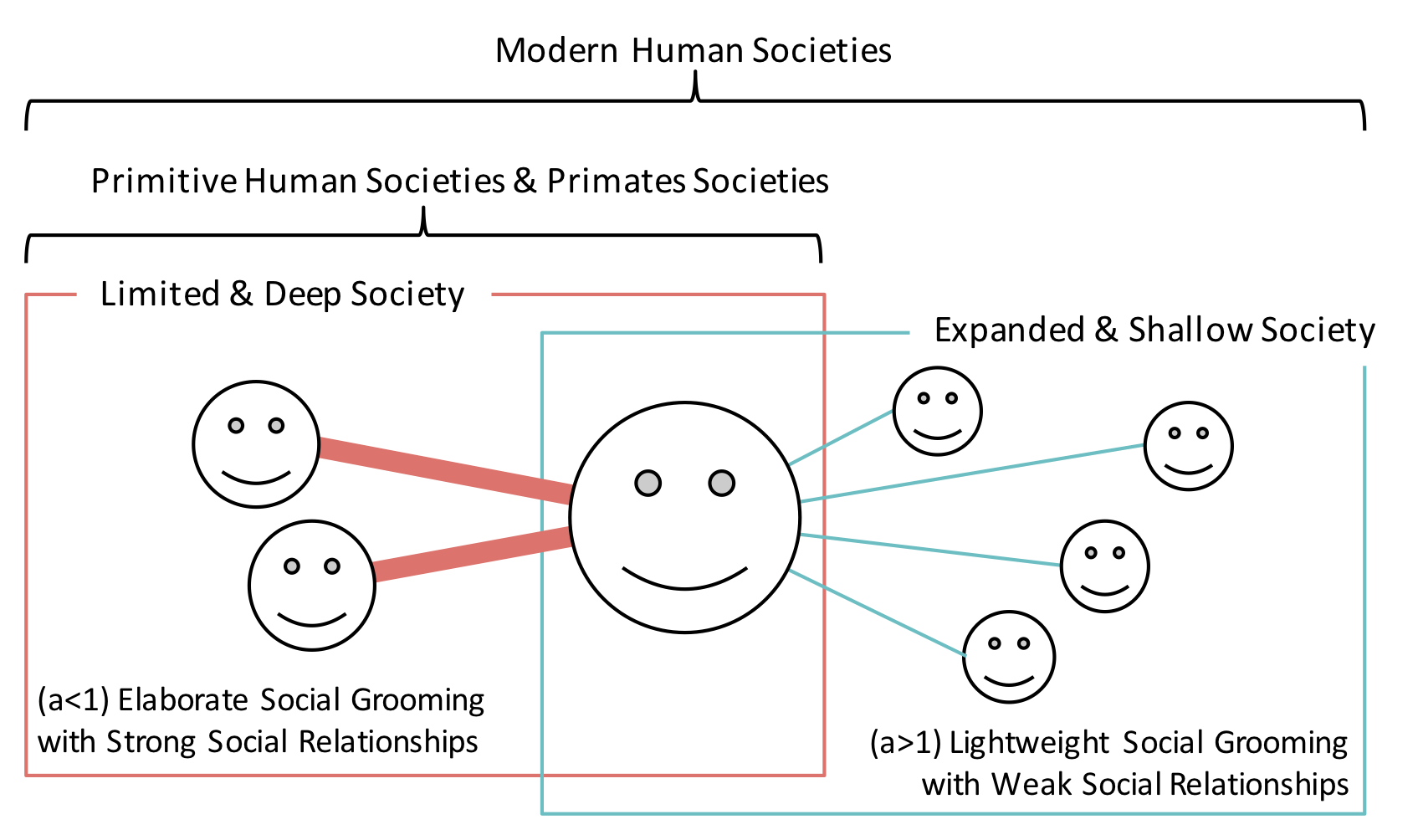}
\caption{
I found two types of social grooming (elaborate social grooming (orange) and lightweight social grooming (green)) and social relationship forms depending on them. People and non-human primates tended to do elaborate social grooming (e.g. face to face communication and fur cleaning in primates) with close social relationships. This social grooming generated limited and deep societies. On the other hand, people tended to do lightweight social grooming (e.g. SNS and E-mail) with weak social relationships. This social grooming generated expanded and shallow societies.
}
\end{center}
\end{figure*}

\begin{figure*}[h!]
\begin{center}
\captionsetup{width=1.4\linewidth}
\includegraphics[width=0.9\linewidth]{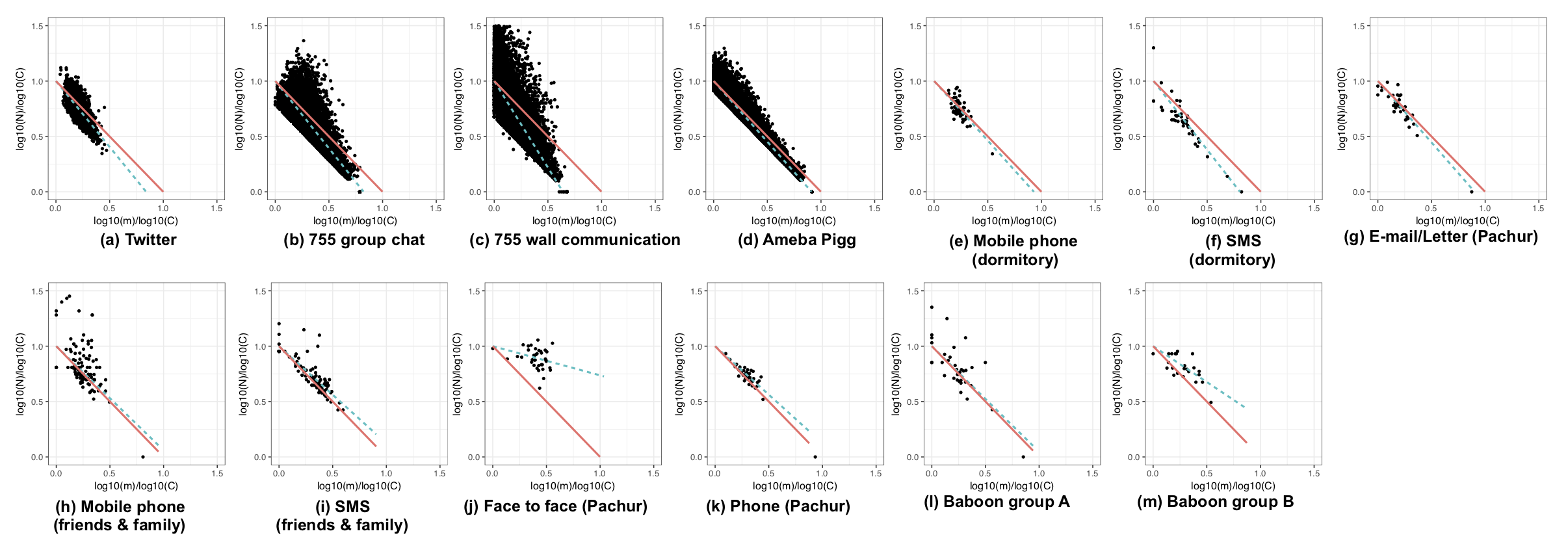}
\caption{
Two types of social relationships separated by the trade-off relationships between $N$ and $m$ (Fig.~a-g are adapted from [9]). This figure shows $\log N/\log C$ and $\log m/ \log C$ to remove the effect of covariate $C$ from the relationships between $N$ and $m$. These were separated by parameter $a$ on $C=Nm^{a}$ model (see Table 1 for details). Fig.~a-g are lightweight social grooming ($a>1$) and Fig.~h-m are elaborate social grooming ($a<1$). The black points show user behaviour data, orange lines show the regression lines of the models when $a=1$ ($C=Nm$), and green dashed lines show the regression lines of $C=Nm^{a}$ models. $a<1$ shows a weak trade-off between $N$ and $m$, as a result, people tended to construct a few strong social relationships. On the other hand, $a>1$ shows a strong trade-off between $N$
and $m$, as a result, people tended to construct many weak social relationships.
}
\end{center}
\end{figure*}

\begin{figure*}[h!]
\begin{center}
\captionsetup{width=1.4\linewidth}
\includegraphics[width=0.9\linewidth]{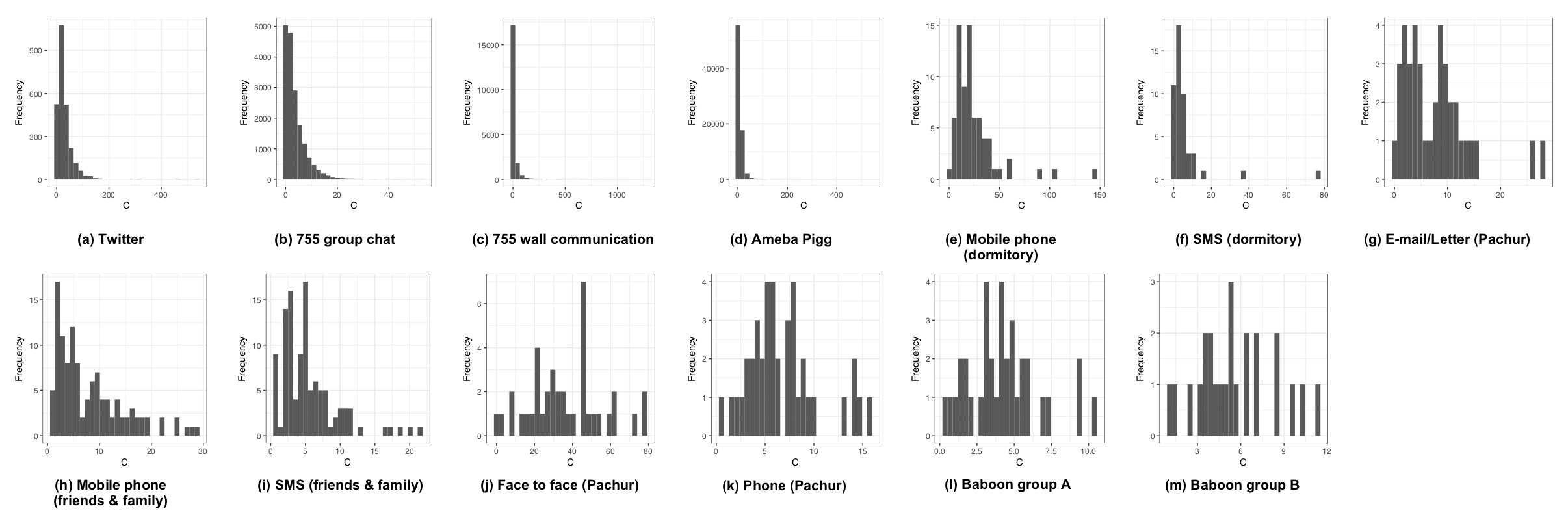}
\caption{
Distributions of total social grooming cost $C$. There is a difference of how frequently individuals used each social grooming method.
}
\end{center}
\end{figure*}

\begin{figure*}[h!]
\begin{center}
\captionsetup{width=1.4\linewidth}
\includegraphics[width=0.9\linewidth]{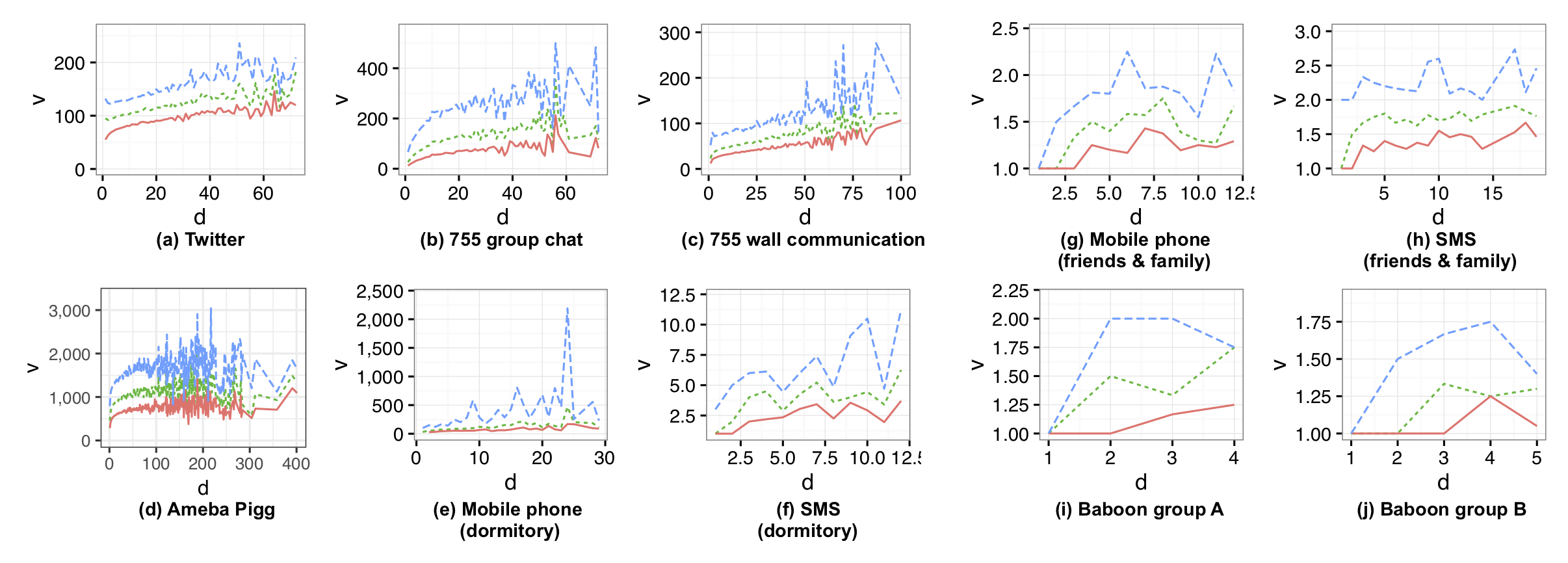}
\caption{
Increasing amount of social grooming per day $v$ by strengths of social relationships $d$  (Fig.~a-h are adapted from [9]). The definitions of an amount of social grooming are shown in Table 2 on the Data-Sets section on ESM. The orange lines are the 25th percentile, the green dotted lines are the 50th percentile and the blue dashed lines are the 75th percentile. These are shown for cases where the number of samples was more than 20 (the ranges of $d$ of Fig.~e-j are short because these were smaller data-sets).
}
\end{center}
\end{figure*}

\begin{figure*}[h!]
\begin{center}
\captionsetup{width=1.4\linewidth}
\includegraphics[width=0.9\linewidth]{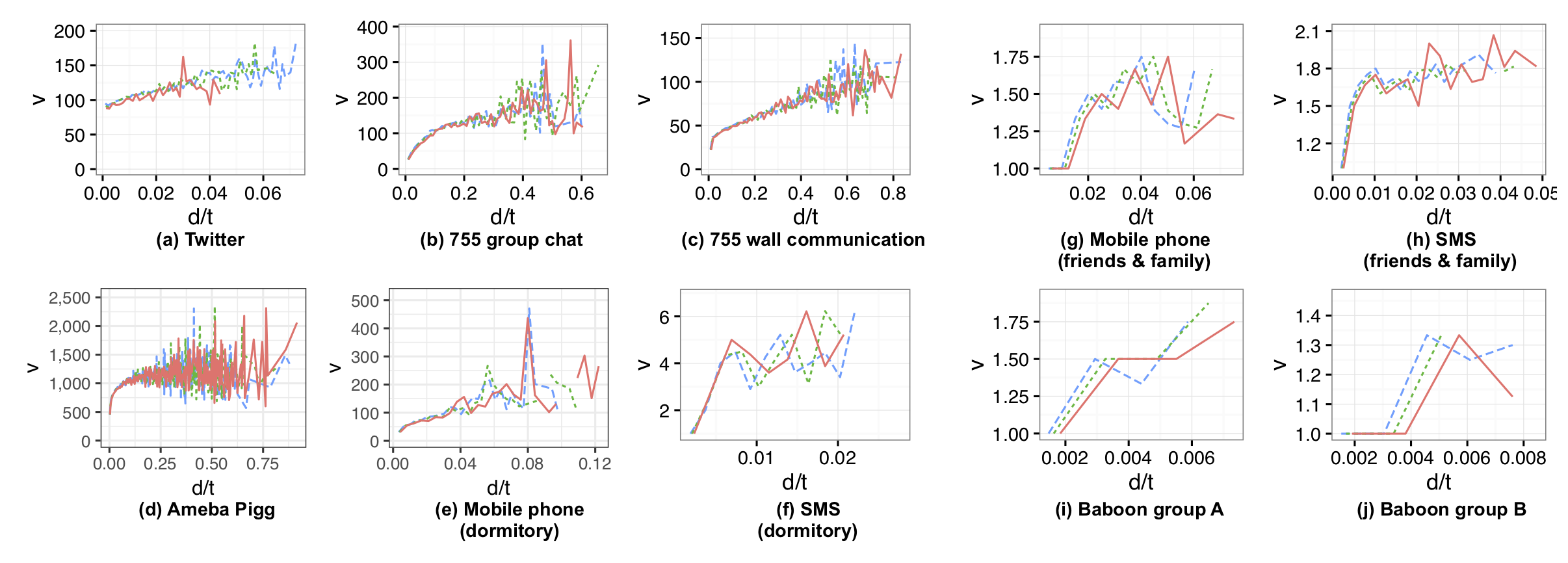}
\caption{
The gradients of the amount of social grooming depending on social grooming density as distinct from those depending on social grooming frequency (Fig.~a-h are adapted from [9]). These figures show a comparison of the medians of $v$ for each social grooming density ($d/t$) for different periods ($t$ is the number of elapsed days). Each line represents entire periods (orange lines), nine-tenths of the periods (green dotted lines), and eight-tenths of the periods (blue dashed lines). These are shown when the number of samples is more than 20 (the ranges of $d$ of Fig.~e-j are short because these were smaller data-sets).
}
\end{center}
\end{figure*}

\begin{figure*}[th!]
\begin{center}
\captionsetup{width=1.4\linewidth}
\includegraphics[width=0.9\linewidth,clip]{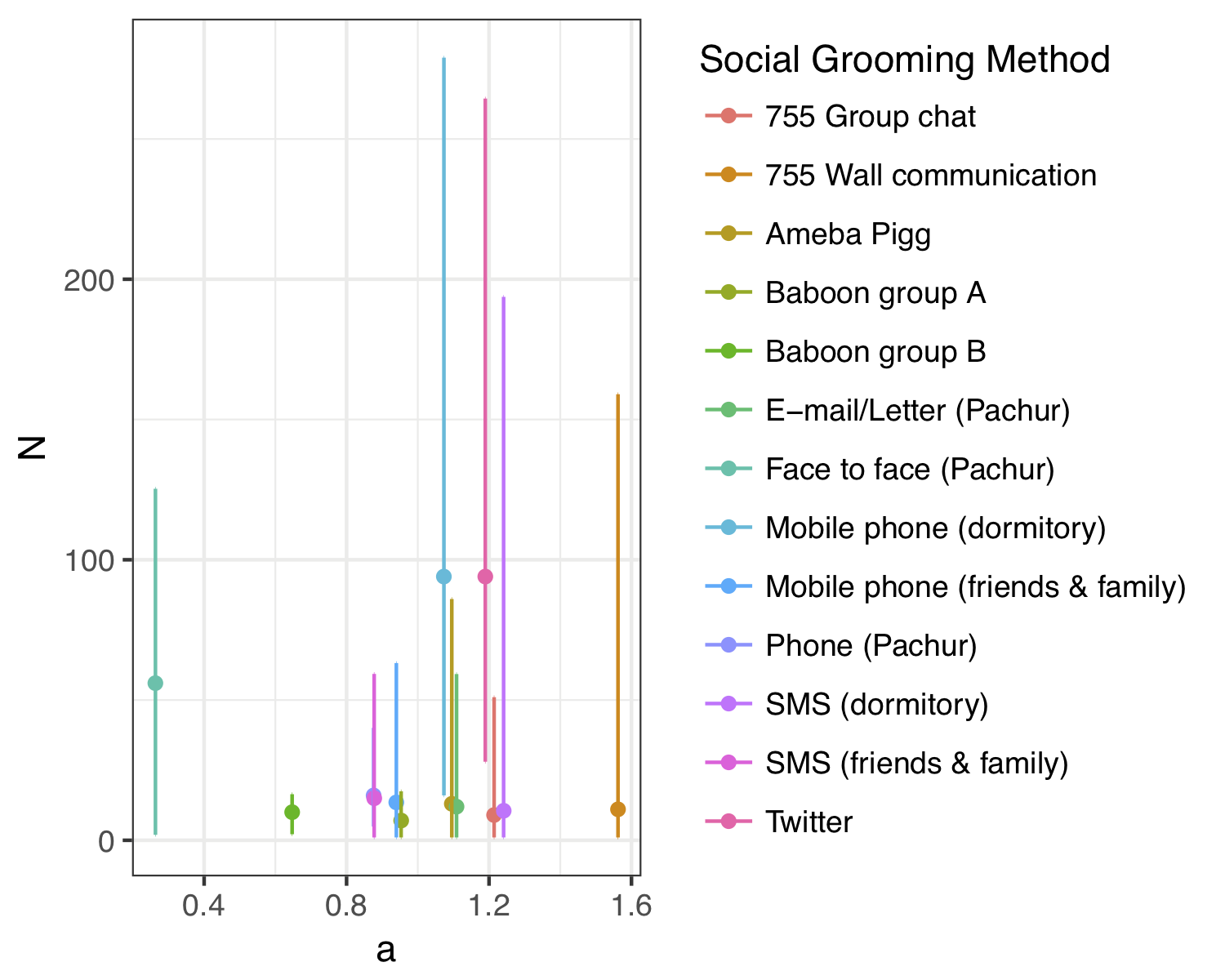}
\caption{
The number of social relationships on each social grooming method. Points show the medians of the number of social relationships. Error bars show ranges from 2.5 percentile to 97.5 percentile. The number of social relationships for each social grooming method did not show clear trends between $a$ and the number of social relationships. This seemed to be caused by the difference in data gathering on the data-sets.}
\end{center}
\end{figure*}

\begin{figure*}[h!]
\begin{center}
\captionsetup{width=1.4\linewidth}
\includegraphics[width=0.9\linewidth]{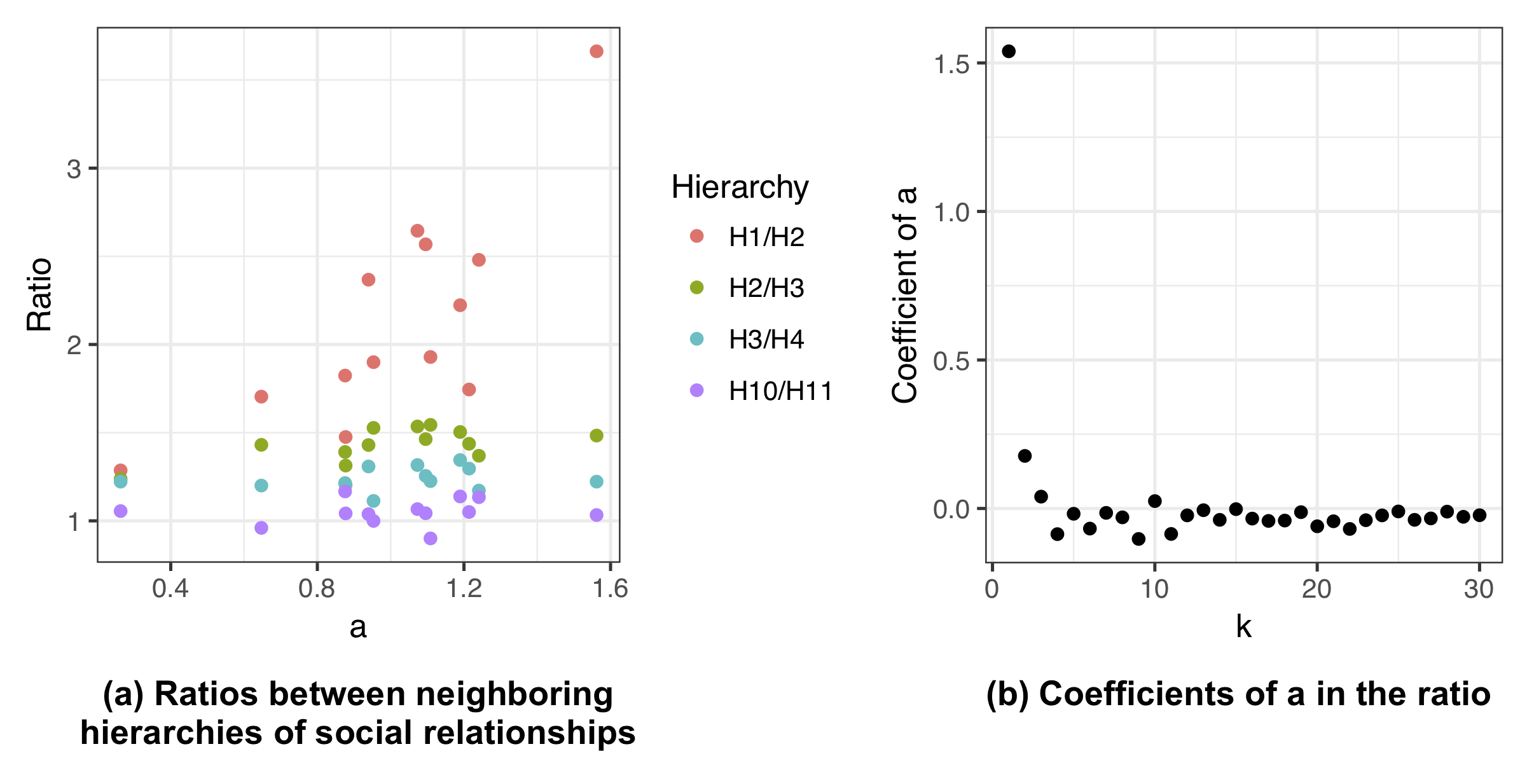}
\caption{
Relationships between $a$ and the size ratio between neighboring hierarchies of social relationships ($H_{k}/H_{k+1}$), where $H_{k}$ is the number of social relationships when $d \geq k$ (i.e., smaller $k$ shows the outer hierarchies).
Fig.~a shows examples of the size ratios ($k=1, 2, 3$, and 10). Fig.~b shows coefficient of a in a regression model ($\beta_{1}$ in  $H_{k}$/$H_{k+1}~Normal(\beta_{1}a+\beta_{0}, \sigma)$) on each $k$ (see ESM Table 1 for details), where the p-values of coefficients of $a$ of $k=1, 2$ were significant (at the 5\% level). That is, these ratios increased with $a$ only when $k$ is small. This suggests that the trade-off parameter $a$ mainly affected very weak social relationships.
}
\end{center}
\end{figure*}

\begin{figure}[th!]
\begin{center}
\captionsetup{width=1.4\linewidth}
\includegraphics[width=0.9\linewidth]{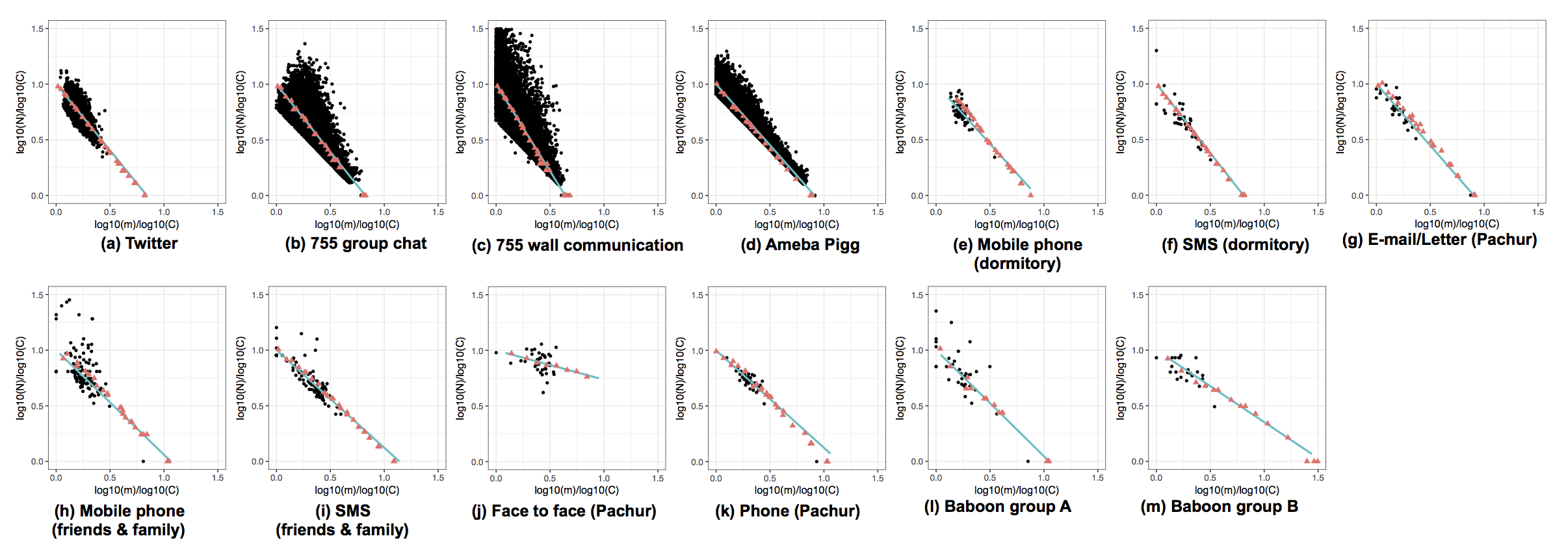}
\caption{
The simulation model fit the data-sets (simulation experiment 1). This shows consistency of the model with two assumptions ($v(w_{ij}) = \alpha w_{ij} + 1$ and $G(a, \alpha; C, m)$). The simulations made fit the results to the regression lines of all data-sets (that is, green and dashed lines in Fig.~2), where these fitting parameters were $\alpha$. Very good fits were observed between the simulation results (orange triangles) and the regression lines (green lines). The parameters $\alpha$ of $v(w_{ij})$ and $G(a, \alpha; C, m)$ were shown in Table 2.
}
\end{center}
\end{figure}

\begin{figure}[th!]
\begin{center}
\captionsetup{width=1.4\linewidth}
\includegraphics[width=0.9\linewidth]{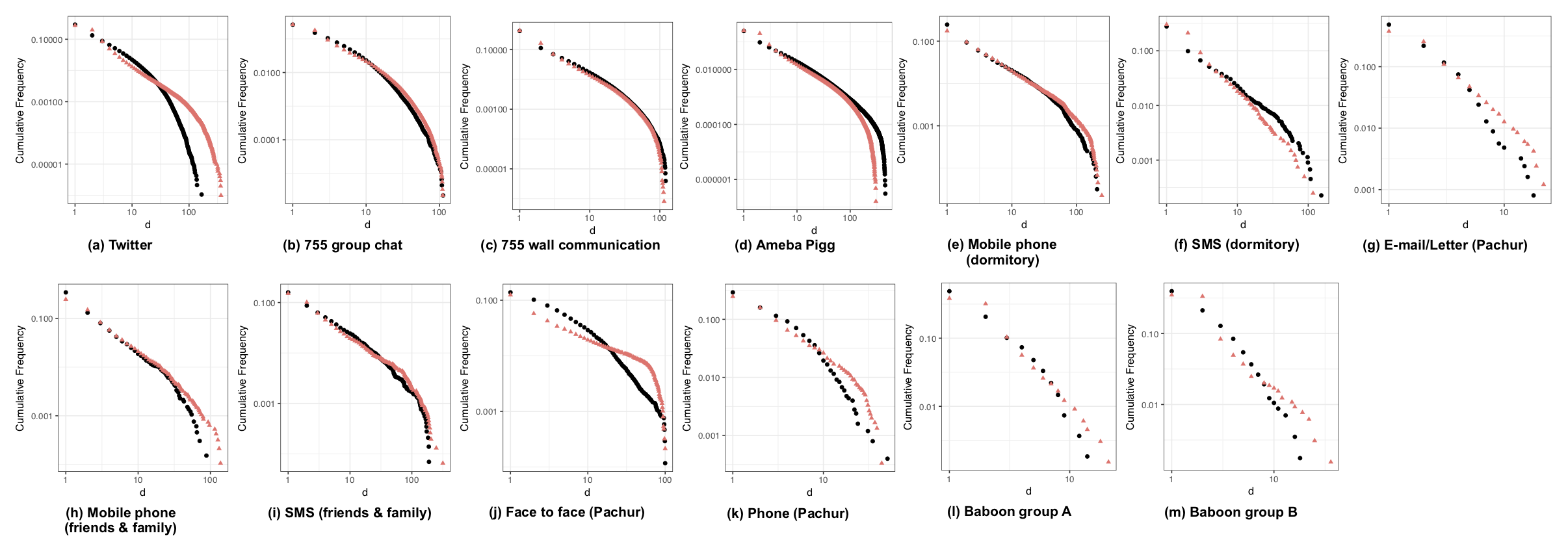}
\caption{
The cumulative distributions of strengths of social relationships $d_{ij}$ of each data-set (black points) and a simulation result of each data-set (orange triangle points) in simulation experiment 1. The simulation results roughly show similar trends with each data-set excluding Face to face (Pachur).
}
\end{center}
\end{figure}

\begin{figure}[th!]
\begin{center}
\captionsetup{width=1.4\linewidth}
\includegraphics[width=0.9\linewidth]{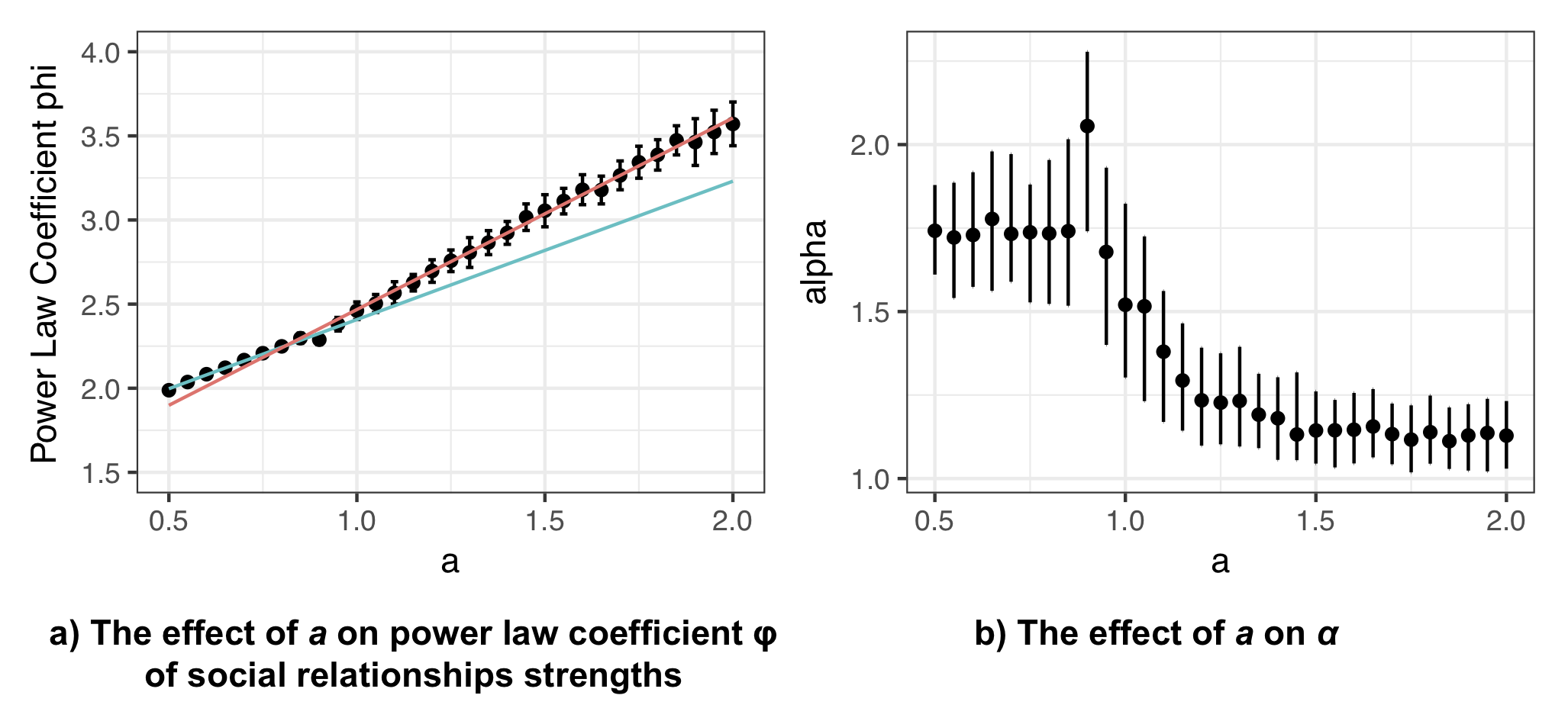}
\caption{
A change of skewed social relationship forms (power law coefficients $\phi$ on distributions of social relationship strengths $d_{ij}$) around a threshold $a=0.8$ (Fig.~a). The black points are the mean $\phi$ in the lowest twenty error value $e_{a\alpha}$ in each $a$ and the error bars are standard deviations of these values. The orange and green lines are the result of the linear regression model with the threshold at $a=0.8$. That is, the green line shows the gradient of coefficients $\phi$ ($\beta_{2}$) in $a \in [0.5, 0.8)$ and the orange line shows the gradient of $\phi$ ($\beta_{1}$) in $a \in [0.8, 2.0]$. The gradient of $\phi$ when $a \geq 0.8$ was larger than when $a<0.8$,
i.e. expanded and shallow social relationship forms. That is, power law coefficients $\phi$ when $a \geq 0.8$ tended to change more than when $a<0.8$. Simultaneously, $\alpha$ was dramatically decreased when $a\geq0.8$, i.e. individuals decreased social grooming time for strong social relationships (Fig.~b). The black points show mean $\alpha$ in the lowest twenty error value $e_{a\alpha}$ in each $a$. The error bars show their standard deviations. These values of the coefficients and $\alpha$ were calculated by the individual-based simulations.}
\end{center}
\end{figure}

\begin{figure}[th!]
\begin{center}
\captionsetup{width=1.4\linewidth}
\includegraphics[width=0.9\linewidth]{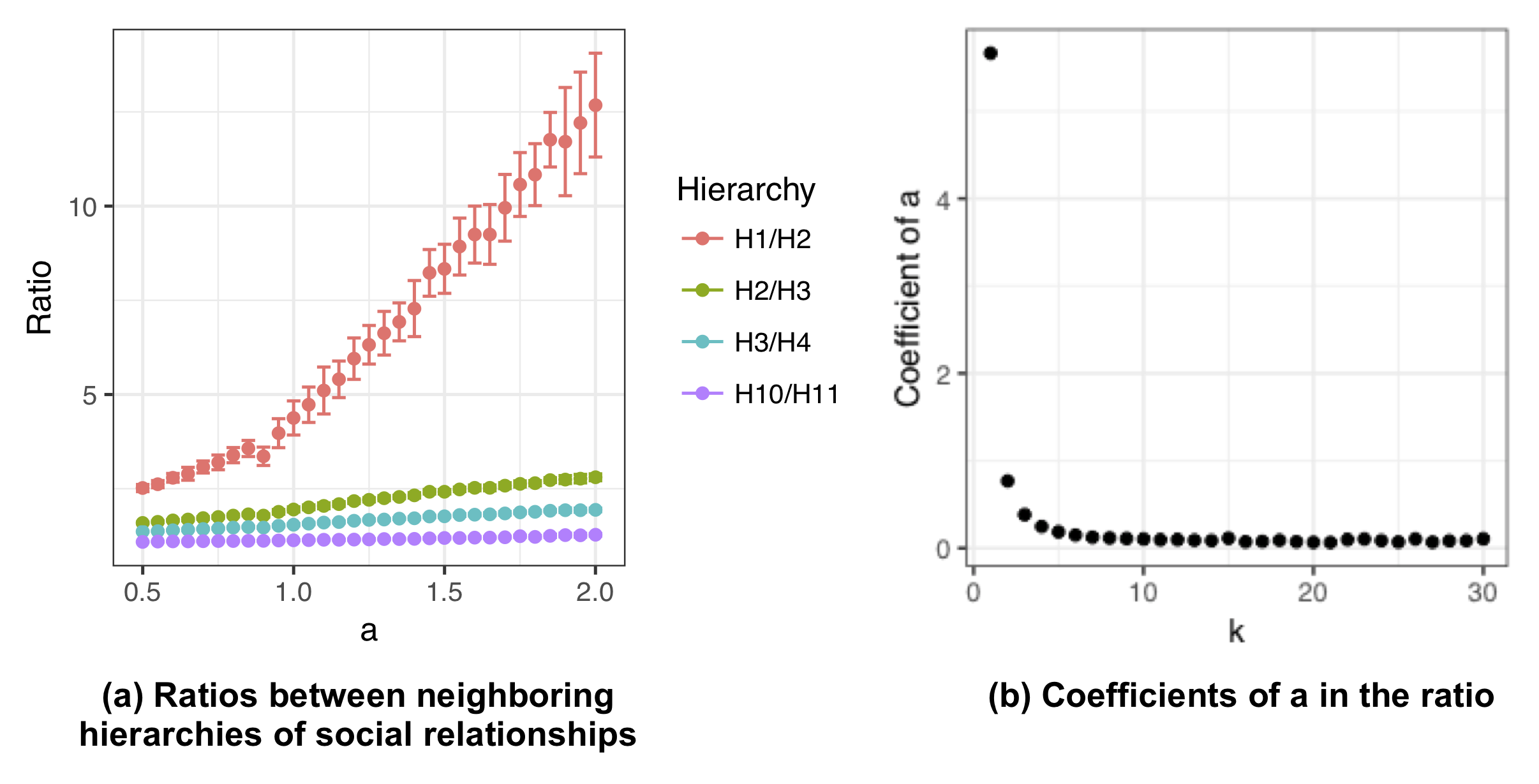}
\caption{
The ratio between neighboring hierarchies ($H_{k}/H_{k+1}$). $H_{k}$ is the number of social relationships when $d > k$. Fig.~a shows examples of the size ratios ($k=1, 2, 3$, and 10). Fig.~b shows coefficients of a in a regression model ($\beta_{1}$ in $H_{k}$/$H_{k+1}~Normal(\beta_{1}a+\beta_{0}, \sigma)$) on each $k$ (see ESM Table 2 for details), where all p-values of coefficients of $a$ were significant (at the 5\% level). These ratios increased with $a$ only when $k$ is very small. The ratio between neighboring hierarchies of very weak social relationships $H_{1}/H_{2}$ significantly changed around $a$=0.8.}
\end{center}
\end{figure}

\begin{figure}[th!]
\begin{center}
\captionsetup{width=1.4\linewidth}
\includegraphics[width=0.9\linewidth]{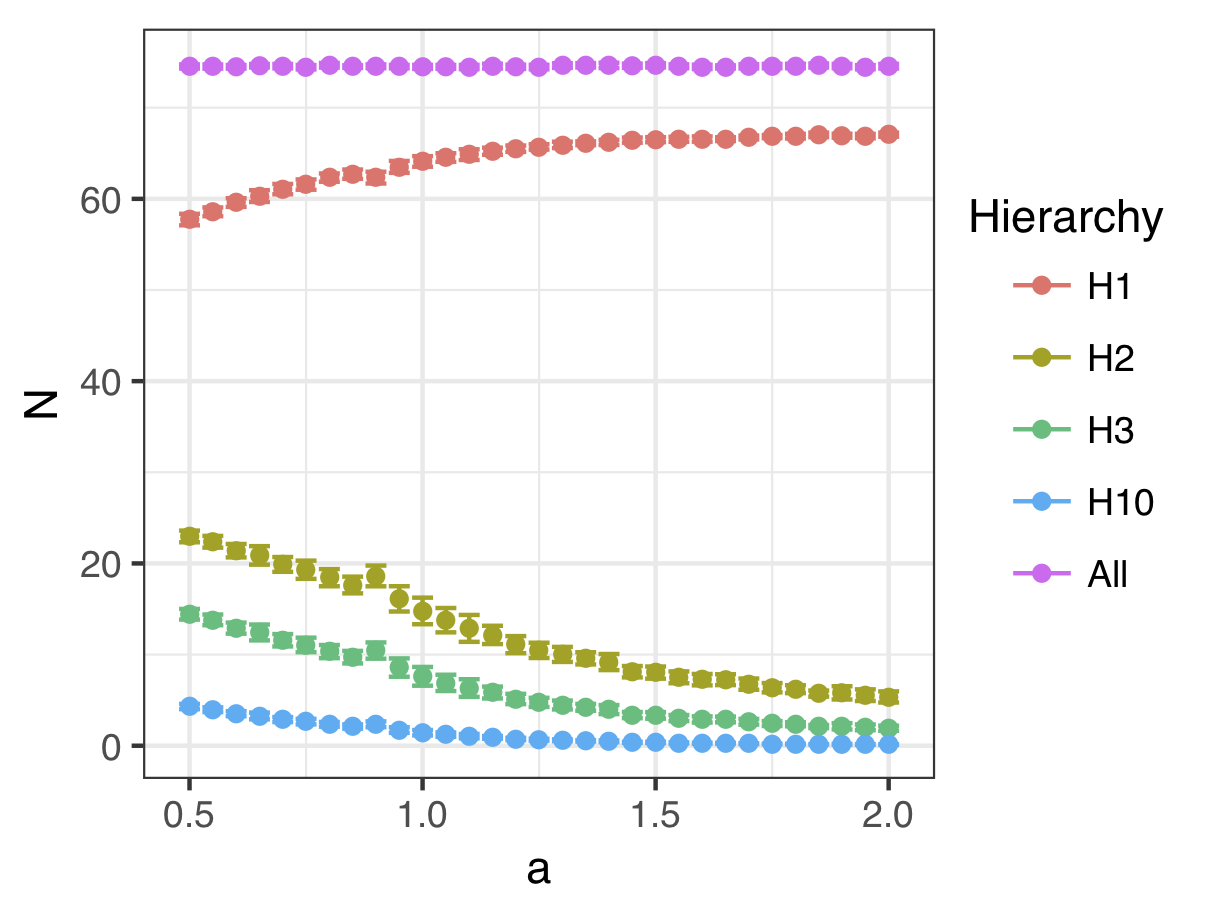}
\caption{
Each hierarchy size depends on $a$. Points show the means of the number of social relationships. Error bars show standard deviation. Hierarchy size $H_{1}$ including very weak social relationships increased with $a$. In contrast, hierarchy sizes $H_{k}$ on $k \geq 2$ decreased with $a$. The number of social relationships (ALL) was approximately constant because this was determined by the settings of the simulation.
}
\end{center}
\end{figure}

\begin{figure}[th!]
\begin{center}
\captionsetup{width=1.4\linewidth}
\includegraphics[width=0.9\linewidth]{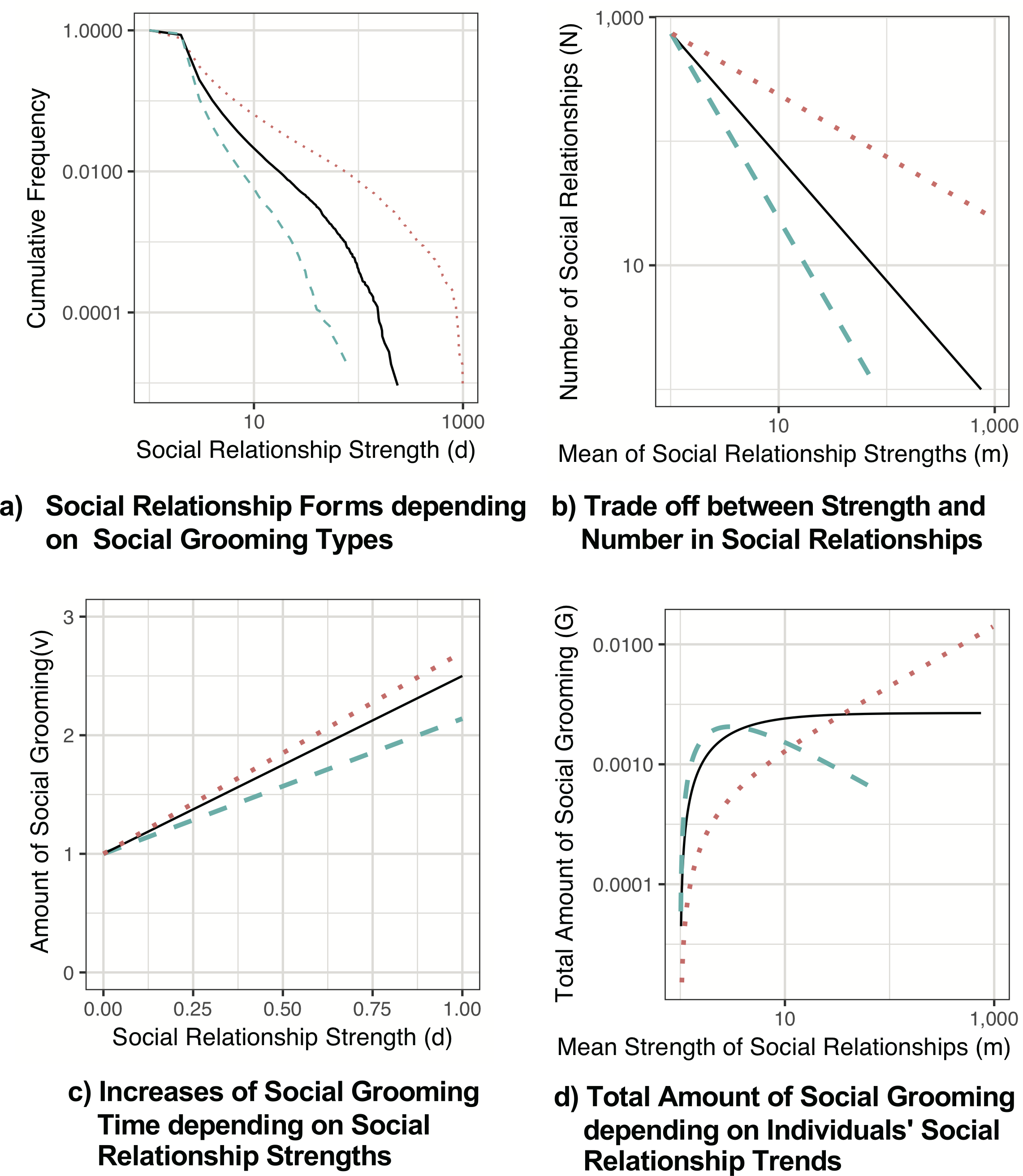}
\caption{
The relationship between social relationship forms (Fig.~a), the trade-off of social grooming methods (Fig.~b), and human social behaviour (Fig.~c and d). In these figures, the orange lines show elaborate social grooming ($a<1$), the green lines show lightweight social grooming ($a>1$), and the black lines are the threshold ($a=1$). Social grooming methods were separated by parameter $a$ of trade-off relationships between the number and strength of social relationships (Fig.~b). People changed their social grooming behaviour depending on $a$. The first was the gradients of an amount of social grooming from a individual to another (the gradients of Fig.~c) increased with each strength of the social relationship. The stronger the social relationships, the greater the amount of social grooming was spent in those relationships. The gradients of lightweight social grooming ($a>1$) were lower than those of elaborate social grooming ($a<1$). The second was the total amount of social grooming to reinforce social relationships $G$ of each individual (Fig.~d). People having a few close social relationships tended to do social grooming frequently when $a<1$ (the amount of social grooming was large). In contrast, people having many weak social relationships tended to do social grooming frequently when $a>1$. The total amount of social grooming is normalized to compare each $a$. These two different behaviours depending on the trade-off changed social relationship forms (Fig.~a). These figures were drawn by using the Twitter data-set with $a=0.5$ (orange), $a=1.0$ (black), and $a=1.5$ (green) in simulation experiment 2. The total amount of social grooming in Fig.~d is normalized to compare each $a$.
}
\end{center}
\end{figure}

\begin{table}[h!]
\captionsetup{width=1.4\linewidth}
  \begin{center}
  \caption{
The results of the regression analysis in Fig.~2.
The t-values and the p-values of $a$ measuring the statistical uncertainty in coefficient $a$ are larger than $1$ when $a > 1$ and the statistical uncertainty in coefficient $a$ are smaller than $1$ when $a < 1$.
The t-values and the p-values of $b$ measuring the statistical uncertainty in coefficient $b$ are not equal to $0$.
The coefficient $a$ was larger than $1$ in lightweight social grooming methods (Twitter, 755 group chat, 755 wall communication, Ameba Pigg, Mobile phone (dormitory), SMS (dormitory), and E-mail/Letter (Pachur)).
On the other hand, the coefficient $a$ was smaller than $1$ in elaborate social grooming methods (Mobile phone (friends \& family), SMS (friends \& family), Face to face (Pachur), Phone (Pachur), Baboon group A, and Baboon group B).
Their adjusted R-squared values were very high.
  }
  \footnotesize
  \hspace*{-0.7cm}
    \begin{tabular}{c|r|c|r|r|r|r}
      \shortstack{Social Grooming \\Method} & \shortstack{Adjusted \\R-squared} & Coef & Estimate & \shortstack{Standard\\ Error} & t-value & p-value \\ \hline \hline
        Twitter & $0.990$ & $a$ & $1.18957$ & $0.02326$ & $51.15$ & $4.4 \times 10^{-16}$ \\
                && $b$ & $1.30935$ & $0.00682$ & $192.12$ & Less than $2.0 \times 10^{-16}$ \\ \hline
        755 Group chat & $0.974$ & $a$ & $1.21423$ & $0.00464$ & $46.17$ & Less than $2.0 \times 10^{-16}$ \\
          && $b$ & $1.26977$ & $0.00229$ & $553.5$ & Less than $2.0 \times 10^{-16}$ \\ \hline
        755 Wall & $0.959$ & $a$ & $1.56214$ & $0.00625$ & $89.94$ & Less than $2.0 \times 10^{-16}$ \\
        communication && $b$ & $1.47639$ & $0.00277$ & $533.2$ & Less than $2.0 \times 10^{-16}$ \\ \hline
        Ameba Pigg & $0.997$ & $a$ & $1.09541$ & $0.00074$ & $128.24$ & Less than $2.0 \times 10^{-16}$  \\
          && $b$ & $1.09395$ & $0.00031$ & $3487$ & Less than $2.0 \times 10^{-16}$ \\ \hline
        Mobile phone & $0.994$ & $a$ & $1.07332$ & $0.15756$ & $0.48$ & $3.2 \times 10^{-1}$ \\
               (dormitory) && $b$ & $1.25628$ & $0.04689$ & $26.795$ & Less than $2.0 \times 10^{-16}$ \\ \hline
        SMS & $0.990$ & $a$ & $1.24089$ & $0.07815$ & $3.08 $ & $3.5 \times 10^{-3}$ \\
        (dormitory) && $b$ & $1.21949$ & $0.02995$ & $40.72$ & Less than $2.0\times 10^{-16}$ \\ \hline
        E-mail/Letter  & $0.988$ & $a$ & $1.10884$ & $0.17744$ & $0.613$ & $2.7 \times 10^{-1}$ \\
        (Pachur) && $b$ & $1.14310$ & $0.04267$ & $26.789$ & Less than $2.0\times 10^{-16}$ \\ \hline
        Mobile phone & $0.978$ & $a$ & $0.93982$ & $0.13896$ & $0.43$ & $3.3 \times 10^{-1}$ \\
        (friends \& family) && $b$ & $1.23703$ & $0.04531$ & $27.300$ & Less than $2.0\times 10^{-16}$ \\ \hline
        SMS & $0.992$ & $a$ & $0.87749$ & $0.05906$ & $2.07 $ & $2.0 \times 10^{-2}$ \\
        (friends \& family) && $b$ & $1.04883$ & $0.02360$ & $44.45$ & Less than $2.0\times 10^{-16}$ \\ \hline
        Face to face & $0.990$ & $a$ & $0.26334$ & $0.18489$ & $3.984$ & $1.6 \times 10^{-4}$ \\
        (Pachur) && $b$ & $1.02148$ & $0.08065$ & $12.666$ & 5.1 $2.0\times 10^{-15}$ \\ \hline
        Phone & $0.997$ & $a$ & $0.87585$ & $0.09219$ & $1.35$ & $9.3 \times 10^{-2}$ \\
        (Pachur) && $b$ & $1.07172$ & $0.03087$ & $34.717$ & Less than $2.0\times 10^{-16}$ \\ \hline
        Baboon group A & $0.985$ & $a$ & $0.95319$ & $0.16301$ & $0.29 $ & $3.9 \times 10^{-1}$ \\
                && $b$ & $1.17293$ & $0.05034$ & $23.30$ & Less than $2.0\times 10^{-16}$ \\ \hline
        Baboon group B & $0.991$ & $a$ & $0.64714$ & $0.13587$ & $2.60$ & $8.2 \times 10^{-3}$ \\
                && $b$ & $1.07354$ & $0.04763$ & $22.539$ & Less than $2.0\times 10^{-16}$ \\
        \end{tabular}
    \end{center}
\end{table}

\begin{table}[th!]
  \begin{center}
  \captionsetup{width=1.4\linewidth}
      \caption{
  The parameters $\alpha$ of $v(w_{ij})$ and $G(a, \alpha; C, m)$ in Simulation Experiment 1.
  }
  \footnotesize
    \begin{tabular}{c|r}
      Communication System & $\alpha$  \\ \hline \hline
      Twitter & 1.034927  \\ \hline
      755 group chat & 1.206970  \\ \hline
      755 wall communication & 1.131248  \\ \hline
      Ameba Pigg & 0.946045  \\ \hline
      Mobile phone (dormitory) & 2.734375  \\ \hline
      SMS (dormitory)& 1.019287 \\ \hline
      E-mail/Letter (Pachur) & 1.708984 \\ \hline
      Mobile phone (friends \& family) & 1.887512 \\ \hline
      SMS (friends \& family) & 1.562500 \\  \hline
      Face to face (Pachur) & 2.148438 \\ \hline
      Phone (Pachur) & 1.660156 \\ \hline
      Baboon group A & 1.044464 \\ \hline
      Baboon group B & 1.020813 \\
       \end{tabular}
    \end{center}
\end{table}

\begin{table}[th!]
  \begin{center}
  \captionsetup{width=1.4\linewidth}
  \caption{High correlations of $\log G(a, \alpha; C_i, m_i)$ and $\log G_i$, where $G_i$ is a summation of $i$'s amount of social grooming per day without initial social grooming with strangers, i.e. the actual amount of social grooming.
  $\alpha$ was decided by simulation experiment 1.
  }
  \footnotesize
  \begin{tabular}{c|r|r}
    Communication System & Correlation & p-value \\ \hline \hline
    Twitter & 0.8399704 & Less than $2.0 \times 10^{-16}$ \\ \hline
    755 group chat & 0.7057249  & Less than $2.0 \times 10^{-16}$ \\ \hline
    755 wall communication & 0.8130153  & Less than $2.0 \times 10^{-16}$ \\ \hline
    Ameba Pigg & 0.7678780 & Less than $2.0 \times 10^{-16}$ \\ \hline
    Mobile phone (dormitory) & 0.7434336  & $5.0\times 10^{-14}$  \\ \hline
    SMS (dormitory)& 0.9079204 & Less than $2.0 \times 10^{-16}$  \\ \hline
    Mobile phone (friends \& family) & 0.9749272 & Less than $2.0 \times 10^{-16}$ \\ \hline
    SMS (friends \& family) & 0.9733798  & Less than $2.0 \times 10^{-16}$ \\  \hline
    Baboon group A & 0.9789251 & Less than $2.0 \times 10^{-16}$  \\ \hline
    Baboon group B & 0.9790673 & Less than $2.0 \times 10^{-16}$\\
     \end{tabular}
    \end{center}
\end{table}

\begin{table}[t!]
\captionsetup{width=1.4\linewidth}
  \caption{
The results of the linear regression model with the threshold at $a=0.8$.
This adjusted R-squared value was $0.978$.
  }
  \begin{center}
  \footnotesize
  \begin{tabular}{c|r|r|r|r}
    Coefficient & Estimate & Standard Error & t-value & p-value \\ \hline \hline
    $\beta_1$ & 0.87282  &  0.07979 & 10.940  & Less than $2.0\times 10^{-16}$ \\ \hline
    $\beta_2$ & 0.27829  &  0.08032  & 3.465  & $5.68\times 10^{-15}$ \\ \hline
    $\beta_3$ & -0.24724  &   0.05208 &  -4.747 & $2.56\times 10^{-6}$ \\ \hline
    $\beta_0$ & 1.55549  &  0.05033 & 30.906  & Less than $2.0\times 10^{-16}$ \\
      \end{tabular}
    \end{center}
\end{table}

\end{document}